\documentclass[reprint,
superscriptaddress,
 amsmath,amssymb,
prb,
]{revtex4-2}

\usepackage{graphicx}% Include figure files
\usepackage{dcolumn}% Align table columns on decimal point
\usepackage{bm}% bold math
\usepackage{comment}
\usepackage{xcolor}
\begin{document}

\preprint{APS/123-QED}

\title{Freezing of the Lattice in the Kagome Lattice Heisenberg Antiferromagnet Zn-barlowite ZnCu$_3$(OD)$_6$FBr}% Force line breaks with \\

\author{Jiaming Wang}
\affiliation{%
 Department of Physics and Astronomy, McMaster University, Hamilton, Ontario, L8S 4M1, Canada
}%
\author{Weishi Yuan}
\affiliation{%
 Department of Physics and Astronomy, McMaster University, Hamilton, Ontario, L8S 4M1, Canada
}%

\author{Philip M.Singer}
\affiliation{Department of Chemical and Biomolecular Engineering, Rice University, Houston, Texas, 77005, USA}%
 
 \author{Rebecca W. Smaha}
 \affiliation{Stanford Institute for Materials and Energy Sciences, SLAC National Accelerator Laboratory, Menlo Park, CA 94025, USA}
 \affiliation{Department of Chemistry, Stanford University, Stanford, CA 94305, USA}

\author{Wei He}
\affiliation{Stanford Institute for Materials and Energy Sciences, SLAC National Accelerator Laboratory, Menlo Park, CA 94025, USA}
\affiliation{Department of Materials Science and Engineering, Stanford University,Stanford, CA 94305, USA}

\author{Jiajia Wen}
\affiliation{Stanford Institute for Materials and Energy Sciences, SLAC National Accelerator Laboratory, Menlo Park, CA 94025, USA}

\author{Young S. Lee}
\affiliation{Stanford Institute for Materials and Energy Sciences, SLAC National Accelerator Laboratory, Menlo Park, CA 94025, USA}
\affiliation{Department of Applied Physics, Stanford University, Stanford, CA 94305, USA}

\author{Takashi Imai}
\affiliation{%
 Department of Physics and Astronomy, McMaster University, Hamilton, Ontario, L8S 4M1, Canada
}%


\date{\today}% It is always \today, today,

\begin{abstract}

{We use $^{79}$Br nuclear quadrupole resonance (NQR) to demonstrate that ultra slow lattice dynamics set in below the temperature scale set by the Cu-Cu super-exchange interaction $J$~($\simeq160$~K) in the kagome lattice Heisenberg antiferromagnet Zn-barlowite.  The lattice completely freezes  below 50~K, and $^{79}$Br NQR lineshapes become twice broader  due to increased lattice distortions.  Moreover, the frozen lattice exhibits an oscillatory component in the transverse spin echo decay, a typical signature of pairing of nuclear spins by indirect nuclear spin-spin interaction.  This indicates that some Br sites form structural dimers via a pair of kagome Cu sites prior to the gradual emergence of spin singlets below $\sim30$~K.  Our findings underscore the significant roles played by subtle structural distortions in determining the nature of the disordered magnetic ground state of the kagome lattice.}

%\begin{description}
%\end{description}
\end{abstract}

\maketitle

%\tableofcontents

Identifying the spin liquid ground state realized in model spin Hamiltonians is the holy grail in the research field of frustrated magnetism \cite{Balents2010,Broholm2020}.  Theoretically, multiple states often compete with each other for the ground state of a given spin Hamiltonian.  This makes theoretical identification of the ground state a non-trivial problem.  Likewise, on the experimental side, each spin liquid candidate material has its own complications, too, often arising from structural disorders.  For example, the non-magnetic interlayer Zn$^{2+}$ sites of the kagome lattice Heisenberg antiferromagnet herbertsmithite ZnCu$_3$(OH)$_6$Cl$_2$ \cite{Shores2005,Helton2007,Freedman2010,Han2012,Imai2008,Olariu2008,Imai2011,Fu2015,Han2016,Zorko2017,Kimchi2018,Khunita2020,WangNatPhys2021,Huang2021,Murayama2021} and Zn-barlowite ZnCu$_3$(OH)$_6$FBr \cite{Feng2017,Smaha2020,Smaha2020_PRM,Tustain2020,WangNatPhys2021,Fu2021,Tustain2021} are occupied by Cu$^{2+}$ defect spins with $\sim15$\% \cite{Freedman2010} and $\sim5$\% \cite{Smaha2020_PRM} probability, respectively.  These defect spins have been generally believed to account for the enhanced magnetic response observed at low temperatures, and mask the intrinsic behavior of the kagome planes.

In addition, recent theoretical works suggest that inhomogeneity in the magnitude of the Cu-Cu super-exchange interaction $J$ ($= 160$~K \cite{Smaha2020} $\sim190$~K\cite{Helton2007}) alone could significantly impact the nature of the ground state, and induce spin singlets with inhomogeneous gaps, accompanied by orphaned localized spins elsewhere within the kagome planes \cite{Kawamura2014,Shimokawa2015,Kawamura2019}.  Moreover, these orphaned spins may account for the enhanced magnetic response at low temperatures \cite{Kawamura2014,Shimokawa2015,Kawamura2019}, even if there are no interlayer Cu$^{2+}$ defect spins or spin vacancies \cite{Singh2010} within the kagome planes.  In fact, our recent $^{63}$Cu nuclear quadrupole resonance (NQR) experiments established that spin singlets gradually emerge with inhomogeneous gaps below $\sim30$~K in both herbertsmithite and Zn-barlowite  \cite{WangNatPhys2021}.  

Motivated by these developments, we explore the structural disorder and their dynamics in Zn-barlowite  ZnCu$_{3}$(OD)$_6$FBr  based on nuclear quadrupole resonance (NQR) at $^{79}$Br sites (nuclear spin 3/2).  The $^{79}$Br NQR frequency $^{79}\nu_{Q}$ and its distribution  probe the local lattice environment and its disorder through the electric field gradient (EFG), while the nuclear spin-lattice $^{79}1/T_1$ and spin-spin $^{79}1/T_2$ relaxation rate shed light on the slow dynamics of the lattice at the timescale set by the inverse of the resonant frequency, $^{79}\nu_{Q}^{-1}\sim0.04~\mu$s.  We will demonstrate that ultra slow lattice dynamics set in below the temperature scale of $J \sim 160$~K, and the lattice freezes below $\sim50$~K with enhanced structural disorder.   Moreover, we will report our discovery of an oscillating component in the spin echo decay curves $M(2\tau) \sim \text{cos}\{\omega_\text{o}(2\tau)\}$ \cite{Hahn1952,Abragam,Kodama2002,Kikuchi2010} induced by indirect nuclear spin-spin interaction $\hbar a_{ij} \hat{\bf{I}}_{i} \cdot \hat{{\bf I}}_{j}$ \cite{Ruderman1954,Slichter,Pennington1989,Pennington1991}, where $\tau$ is the separation time between 90 and 180 degree radio frequency pulses, $M(2\tau)$ is the spin echo amplitude at time $2\tau$, $\hat{{\bf I}}_{i}$ and $\hat{\bf{I}}_{j}$ are the nuclear spin operator at the i-th and j-th site, and $a_{ij}$($\sim \omega_\text{o}$) is the indirect nuclear spin-spin coupling.  In short, the spin echo amplitude oscillates, because nuclear spin $\hat{\bf{I}}_{i}$ precesses about the hyperfine magnetic field generated by nuclear spin $\hat{\bf{I}}_{j}$, and vice versa.    Such oscillations are a typical NMR signature of the pairing of atoms in molecules \cite{Hahn1952} and solids, including Cu-Cu spin singlet dimers in SrCu$_2$(BO$_3$)$_2$ \cite{Kodama2002} and Cu$_2$Sc$_2$Mg$_4$O$_{13}$ \cite{Kikuchi2010}, but unexpected for the kagome lattice in Zn-barlowite.  Our finding indicates that some $^{79}$Br sites  in the frozen lattice form structural dimers encompassing a pair of kagome Cu spin singlets.  The existence of the oscillation with a well defined frequency contrasts with the Gaussian form of spin echo decay $M(2\tau)$ observed for the two-leg Heisenberg ladder in SrCu$_2$O$_3$ \cite{Ishida1996} and Sr$_{14}$Cu$_{24}$O$_{41}$ \cite{Takigawa1998}, in which spin singlets are entangled along the legs in the ladder.  We will explain that the spin echo amplitude oscillation can be used as a probe of entanglement between spin singlets.

\begin{figure}
\centering
\includegraphics[width=3in]{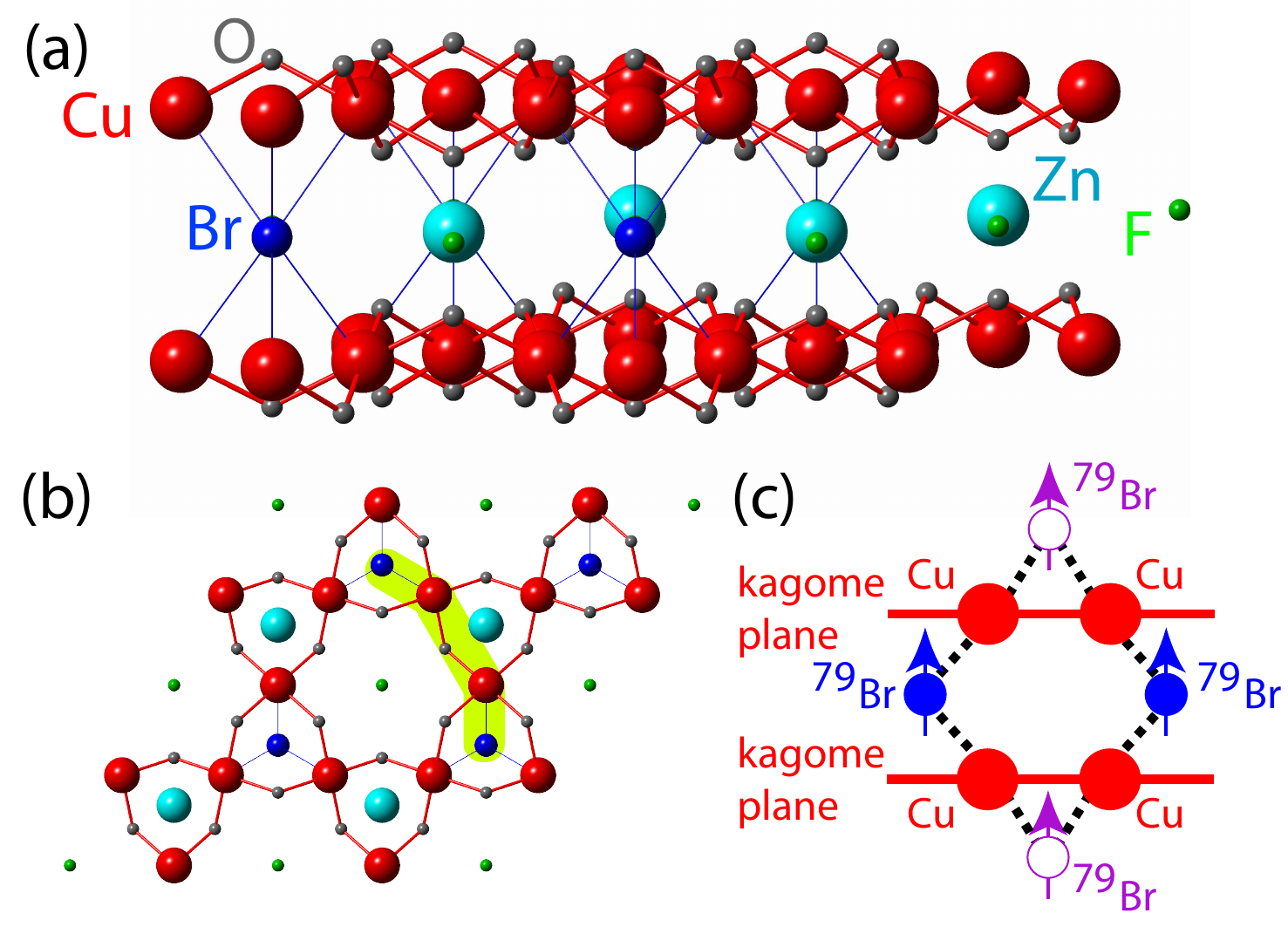}
\caption{(a) The structure of ZnCu$_3$(OD)$_6$FBr.  The kagome layers consist of corner sharing triangles of Cu$^{2+}$ ions with spin-1/2, and stacked on top of each other.  For clarity, D$^{+}$ ions attached to O$^{2-}$ are not shown.  Zn$^{2+}$, F$^{-}$, and Br$^{-}$ form the interlayer.  (b) The c-axis view of the kagome layer, and the interlayer underneath it.  The light green shade schematically represents dimerized Br sites via a pair of Cu sites.  (c) Side view of the kagome planes and $^{79}$Br sites in three interlayers.  Each $^{79}$Br nuclear spin can couple with six nearest-neighbor $^{79}$Br sites within the same interlayer (blue-blue combination) and additional six neighboring $^{79}$Br sites in two adjacent kagome planes (blue-purple combination).  Dashed lines represent the hyperfine coupling between $^{79}$Br nuclear spin and Cu electron spin-1/2.
}
\label{crystal}
\end{figure}

In Fig.1 and 2(a-b), we present the crystal structure of Zn-barlowite and representative $^{79}$Br NQR lineshapes.  We refer readers to \cite{WangNatPhys2021} and Supplemental Materials \cite{SuppMat} for the entire $^{79,81}$Br and $^{63,65}$Cu NQR lineshapes.    We summarize the temperature dependence of the main peak frequency $^{79}\nu_{Q}^{Main}$ in Fig.3(a).  Above 100~K, we found $^{79}\nu_{Q}^{Main} \gtrsim 28.8$~MHz, accompanied by two additional small humps at $^{79}\nu_{Q}^{A} \simeq 28.2$~MHz and $^{79}\nu_{Q}^{B} \simeq 29.4$~MHz  \cite{SuppMat}.  NQR is a local probe, and this indicates that at least three slightly different structural environments exist for $^{79}$Br sites.  We previously identified three sets of $^{2}$D \cite{Imai2011} and $^{17}$O \cite{Fu2015} NMR signals for deuterated herbertsmithite as the main, the nearest neighbor (nn), and the twice more abundant next nearest neighbor sites of the interlayer Cu$^{2+}$ defects occupying the Zn$^{2+}$ sites.  In analogy, we tentatively assign the small hump A and more prominent hump B as the nn and nnn sites, respectively.

\begin{figure}
\centering
\includegraphics[width=3.2in]{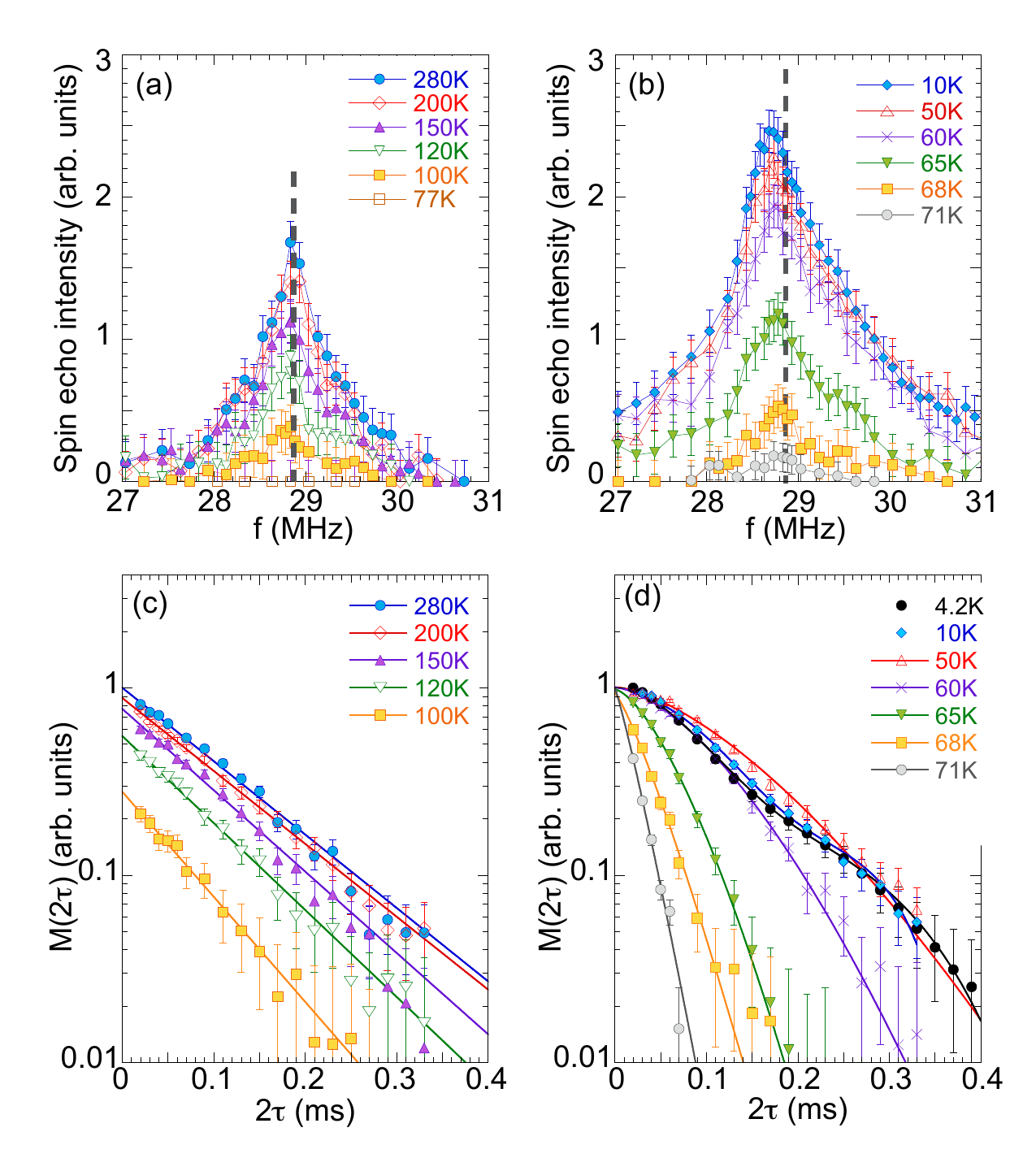}
\caption{Top: representative $^{79}$Br NQR lineshapes observed (a) above and (b) below 75~K measured with a fixed $\tau=30$~$\mu$s, corrected for the Boltzmann factor.  The dashed vertical line marks the peak frequency $^{79}\nu_{Q}^{Main} = 28.85$~MHz at 280~K.  Bottom: spin echo decay curves $M(2\tau)$ observed (c) above and (d) below 75~K.  The fit in (c) is with $M(2\tau) = M_\text{o}\text{exp}\{-(2\tau/T_{2})^{\beta}\}$ with a fixed $\beta = 1$, while $\beta$($\sim1.3$) was allowed to vary above 50~K in (d).  The solid curves through the data below 50~K represent the best two component fit with an oscillatory term,  $M(2\tau) = M_\text{o}[F~\text{cos}\{\omega(2\tau)\}~\text{exp}(-2\tau/D)+(1-F)\text{exp}(-(2\tau/T_{2})^{\beta}$)].   The overall intensity of $M(2\tau)$ in both (c) and (d) is corrected for the Boltzmann factor, and normalized to the $M(2\tau = 0)$ value observed at 280~K.  
}
\label{crystal}
\end{figure}

\begin{figure}
\centering
\includegraphics[width=3.5in]{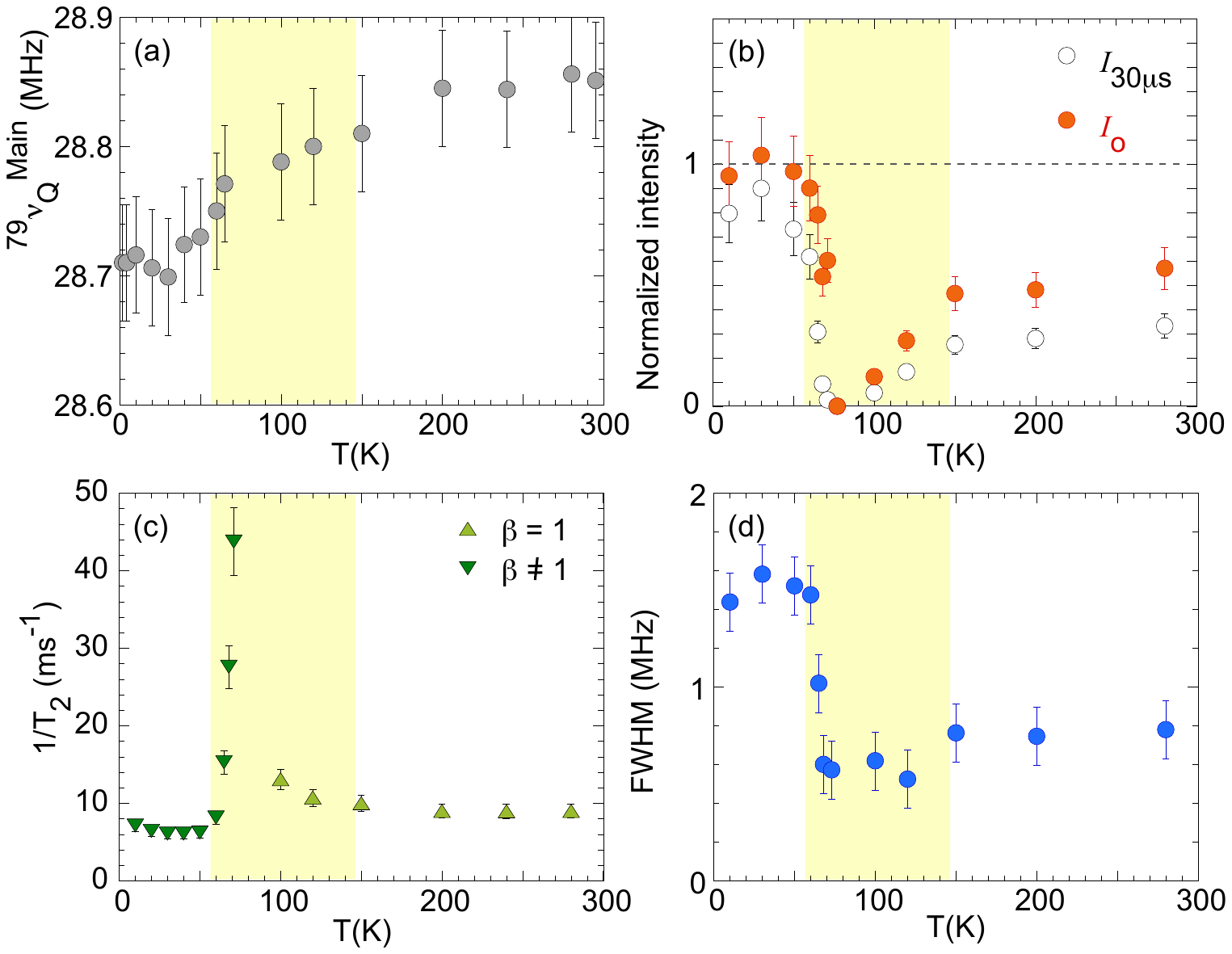}
\caption{(a) The main NQR peak frequency  $^{79}\nu_{Q}^{Main}$.  (b) ($\circ$) The bare integrated intensity $I_{30\mu s}$ of the lineshapes in Fig.\ 2(a-b), and (\textcolor{orange}{$\bullet$}) $I_\text{o}$ at the limit of $2\tau=0$, both corrected for the Boltzmann factor by multiplying temperature $T$.  (c) $1/T_2$ determined from the fit in Fig.2(c-d) with (\textcolor{olive}{$\blacktriangle$}) or without (\textcolor{green}{$\blacktriangledown$}) fixing $\beta = 1$.  (d) FWHM of the $^{79}$Br NQR lineshapes.  We observe only a limited number of $^{79}$Br nuclear spins due to lattice freezing in the yellow shaded region of Figs.~3 and 4.  
}
\label{crystal}
\end{figure}

In Fig.3(b), we summarize the temperature dependence of the intensity $I_{30\mu s}$ of the $^{79}$Br NQR lineshape integrated between 27~MHz and 31~MHz,  measured with a fixed pulse separation time $\tau=30$~$\mu$s.  Also plotted in Fig.~3(b) are the integrated intensity $I_\text{o}$ in the limit of  $2\tau =0$~$\mu$s, estimated from the extrapolation of the transverse spin echo decay curves $M(2\tau)$ at the main peak presented in Fig.2 (c)-(d).  The temperature dependence of $^{79}1/T_2$ determined from $M(2\tau)$ at  $^{79}\nu_{Q}^{Main}$ is summarized in Fig.3(c).  We confirmed that $^{79}1/T_1$ and $^{79}1/T_2$ measured at the hump A and B are comparable to the main peak's at 200~K.  

The $^{79}$Br NQR signals are gradually wiped out below $\sim150$~K.  The signals begin to reemerge below 75~K, followed by quick saturation at $\sim50$~K as $^{79}1/T_{2}$ slows down.  Notice that the main peak intensity $M(2\tau=0)$ extrapolated to $2\tau=0$ is conserved between $\sim280$~K and below 50~K as shown in Fig.\ 2(c-d), but the integrated intensity above 200~K in Fig.\ 3(b) is too small by a factor of $\sim2$ compared with below 50~K. 

Comparison of the $^{79}$Br NQR lineshapes in Fig.2(a-b) reveals two changes across 75~K.  First, the main peak frequency decreases noticeably when the signal intensity is fully recovered below 50~K, as summarized in Fig.~3(a).  Second, the $^{79}$Br NQR lineshapes, already broad at higher temperatures, become nearly twice as broad below 50~K, as summarized in Fig.~3(d).  Analogous lineshape and intensity anomalies are commonly observed when spin freezing takes place in disordered magnetic materials \cite{Hunt2001}.  But there is no evidence for anomalies in spin degrees of freedom around 75~K in $^{19}$F NMR \cite{WangNatPhys2021} and $\mu$SR experiments \cite{Tustain2020}.   Therefore, the observed NQR anomalies must be attributed to the EFG, and we conclude that the structural environments at $^{79}$Br sites become somewhat different and more disordered below $\sim75$~K.  We note that the spatially averaged crystal structure observed by diffraction techniques maintains the perfect kagome symmetry
down to 3~K by neutron powder diffraction and 13~K by synchrotron x-ray diffraction \cite{Smaha2020}.  Herbertsmithite also experiences a structural distortion around 50~K \cite{Imai2008,Fu2015,Zorko2017}, and the interlayer Cu$^{2+}$ defects occupying the Zn$^{2+}$ sites may be causing it, because the $\nu_{Q}$ tensor at the nn $^{17}$O sites changes \cite{Fu2015}.  It remains to be seen if the $\sim5$\% interlayer Cu$^{2+}$ defects play a role in the freezing of lattice distortion in the present case.  Interestingly, the $^{79}$Br NQR lineshapes observed below 60~K for pure barlowite Cu$_4$(OH)$_6$FBr  are very similar \cite{Ranjith2018}.  The NQR results discussed so far do not provide information on the nature of the local structural changes across 75~K, but an important clue is in the shape of the spin echo decay curve $M(2\tau)$.  We will come back to this point below.  

In order to understand the mechanism behind these NQR anomalies across $\sim75$~K, we measured $^{79}1/T_{1}$ at the main peak between 60~K and 280~K.  We also measured $^{81}1/T_{1}$ at $^{81}\nu_{Q} \simeq 24.1$~MHz for the $^{81}$Br sites.  The $^{79,81}1/T_{1}$ results below 60~K were adopted from \cite{WangNatPhys2021}.  For simplicity, we deduced $1/T_1$ by fitting the nuclear spin recovery curve with the conventional stretched exponential, but more elaborate analysis based on the inverse Laplace transform (ILT) \cite{SuppMat,mitchell:PNMRS2012,Singer2020} leads us to the same conclusions; see Supplemental Materials \cite{SuppMat} for the details about the ILTT$_{1}$ analysis technique and related issues, including \cite{mitchell:PNMRS2012,Singer2020,Arsenault2020,Takahashi2019,WangPRB2021,JohnstonPRL2005,Andrew1961,Narath1967}.  We compare $^{79}1/T_{1}$ and $^{81}1/T_{1}$ in Fig.~4(a), and summarize their ratio $R =(^{81}1/T_{1})/(^{79}1/T_{1})$ in Fig.~4(b).  In general, $1/T_1$ measured by NQR for nuclear spin 3/2 may be expressed as $1/T_{1} = 1/T_{1}^{spin} + 1/T_{1}^{lattice}$, where $1/T_{1}^{spin}$ is the magnetic contribution by spin fluctuations, whereas $1/T_{1}^{lattice}$ is caused by lattice fluctuations through the EFG.  In addition,$^{79,81}1/T_{1}^{spin}$ is proportional to the square of the nuclear gyromagnetic ratio $^{79,81}\gamma_{n}$, while $^{79,81}1/T_{1}^{lattice}$ is proportional to the square of the nuclear quadrupole moment $^{79,81}Q$, where $(^{81}\gamma_{n}/^{79}\gamma_{n})^{2} = 1.161$ and $(^{81}Q/^{79}Q)^{2} = 0.698$.  Therefore the ratio $R\simeq 1.161$ observed below 50~K indicates that  $^{79,81}1/T_{1}$ is dominated entirely by Cu spin fluctuations, but additional contributions from lattice fluctuations at the NQR frequency reduce $R$ above 60~K.

\begin{figure}
\centering
\includegraphics[width=3in]{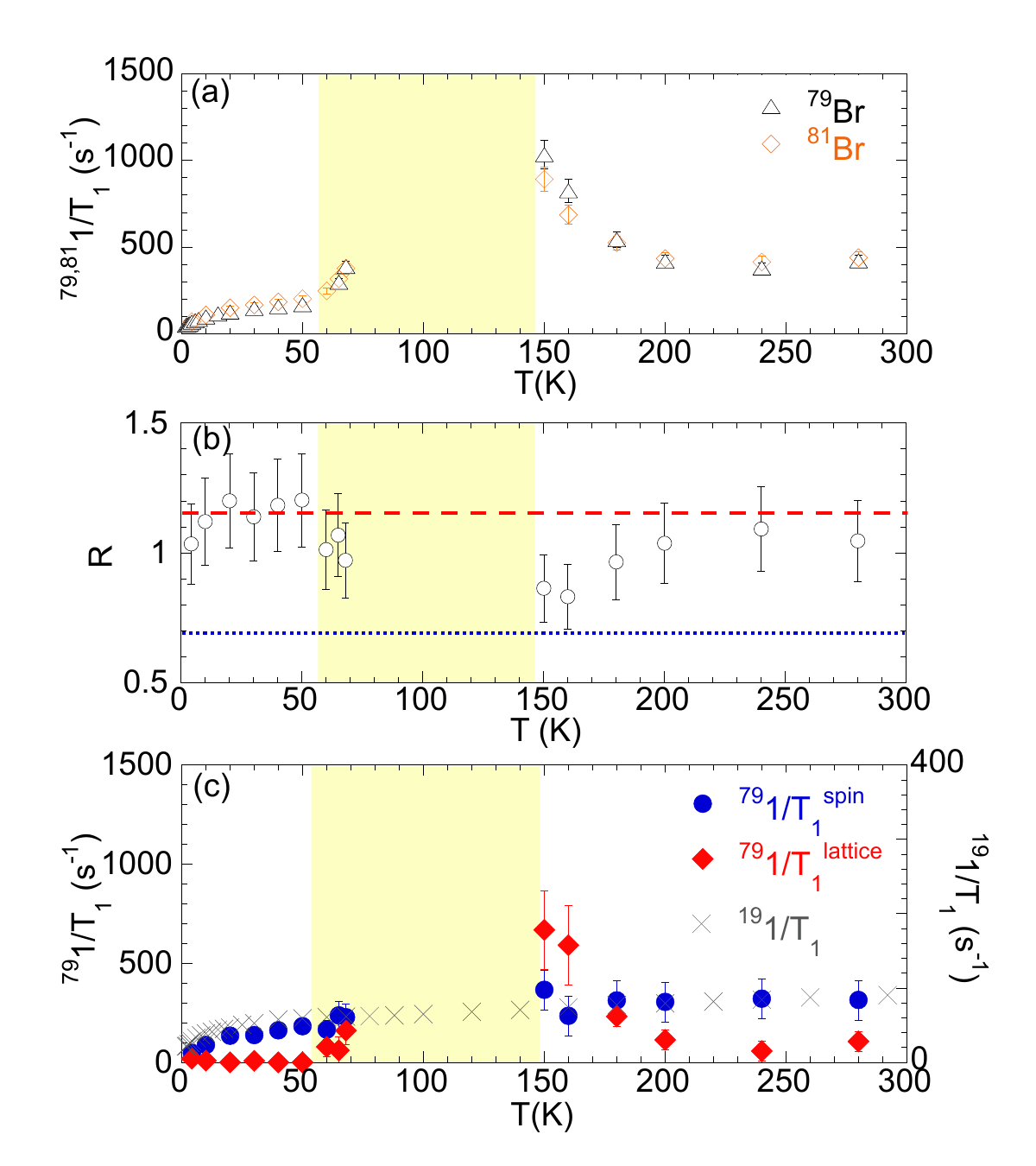}
\caption{(a) $^{79,81}1/T_{1}$ measured at the main peak.  The results below 60~K were adopted from \cite{WangNatPhys2021}.    (b) The isotope ratio $R = (^ {81}1/T_{1})/(^{79}1/T_{1}$).  Dashed and dotted horizontal lines mark $R = 1.161$ (dominated by spin fluctuations)  and $R = 0.698$ (dominated by lattice fluctuations).  (c) $^{79}1/T_{1}^{spin}$ and $^{79}1/T_{1}^{lattice}$ estimated from the results in (a).  Also plotted using the right axis is $^{19}1/T_{1}$ measured at $^{19}$F sites \cite{WangNatPhys2021}.
}
\label{crystal}
\end{figure}

In Fig.~4(c), we estimate $^{79}1/T_{1}^{spin}$ and $^{79}1/T_{1}^{lattice}$ separately by inserting the experimentally observed values of $^{79,81}1/T_{1}$ into $^{79}1/T_{1} = ^{79}1/T_{1}^{spin} + ^{79}1/T_{1}^{lattice}$ and $^{81}1/T_{1} = 1.161(^{79}1/T_{1}^{spin}) + 0.698(^{79}1/T_{1}^{lattice}$).  For comparison, we also present $^{79}1/T_{1}^{spin}$ measured at the $^{19}$F sites \cite{WangNatPhys2021}.  $^{19}$F has nuclear spin 1/2 and lacks nuclear quadrupole moment, and probes only spin fluctuations with no influence of the EFG.  The similarity in the observed temperature dependence between $^{79}1/T_{1}^{spin}$ and $^{19}1/T_{1}$ assures us that our procedures to separate $^{79}1/T_{1}$ into $^{79}1/T_{1}^{spin}$ and $^{79}1/T_{1}^{lattice}$ are working well.

One of the key findings of the present work is that $^{79}1/T_{1}^{lattice}$ undergoes a drastic enhancement below the temperature scale set by $J\simeq 160$~K.  Intuitively, this is easily understandable.  When the temperature is lowered below $J$, neighboring Cu sites become magnetically frustrated, because three Cu spin-1/2's located at the corners of each triangle cannot form singlets all at once.  The effects of this magnetic frustration can be partially alleviated if the lattice distorts and two sites form a dimer at the cost of enhanced elastic energy.  Combined with the aforementioned changes observed for $^{79}\nu_{Q}$ and its distributions, the $^{79}1/T_{1}^{lattice}$ results therefore suggest that magnetic frustration effects play a role in enhancing structural distortions below temperature $\sim J$ through the magneto-elastic coupling effects \cite{Li2020}.  The NQR signal is completely wiped out around 75~K, when the spectral weight of the EFG fluctuations becomes very large around the NQR frequency and enhance $^{79}1/T_1$ and $^{79}1/T_2$;  the NQR signals reemerge below 75~K, because the EFG fluctuations become slower than the NQR frequency.  Once the EFG becomes completely static below 50~K at the NQR measurement time scale of $\sim0.04 \mu$s, the NQR intensity saturates and $^{79}1/T_{1}^{lattice}$ vanishes.   We also emphasize that 100\% of the sample volume is affected by these EFG anomalies, which is why the entire $^{79}$Br NQR signal intensity is wiped out.   

The $T_2$ results summarized in Figs.~2(c-d) and 3(c) provide additional support for the physical picture described in the previous paragraph.  Notice that the spin echo decay curve $M(2\tau)$ below 50~K exhibits a typical Gaussian-Lorentzian form with a negative curvature below $2\tau \sim 0.1$~ms.  This is typical for solids, and consistent with the frozen state of the lattice.  But  $^{79}1/T_2$ is strongly enhanced above 60~K, and the $M(2\tau)$ curves become almost Lorentzian (i.e. exponential).  This is consistent with the motional narrowing effects \cite{Abragam,Slichter} induced by the slowly fluctuating EFG.  The spin echo decay is also nearly Lorentzian above 100~K up to 280~K, hinting the possibly dynamic nature of the lattice even above $T\sim J$.  It might also explain why the integrated intensity above 200~K is too small by a factor of $\sim 2$.

A striking aspect of Fig.~2(d) is that $M(2\tau)$ develops a damped oscillatory component $F \text{cos}[\omega_\text{o} (2\tau)]{\text exp}(-2\tau/D)$ in the frozen state below 50~K, preceding the gradual emergence of Cu-Cu spin singlets below $\sim30$~K \cite{WangNatPhys2021}.  Here, $\omega_\text{o} \simeq 20$\ ms$^{-1}$ and $D \simeq 0.1$\ ms represent the oscillation frequency and damping time constant, respectively, and the fraction of the oscillatory component reaches $F \sim 0.25$ at 4.2~K.   

Analogous oscillatory behaviors were previously reported for spin singlets in dimerized materials SrCu$_2$(BO$_3$)$_2$ \cite{Kodama2002} and Cu$_2$Sc$_2$Mg$_4$O$_{13}$ \cite{Kikuchi2010} with the oscillation frequency $\omega_\text{o} = A_{hf}^{2}/4J$ set by the intra-dimer super exchange $J$ \cite{Pennington1989} ($A_{hf}$ represents the hyperfine coupling between the observed nuclear spin and Cu electron spin, which is unknown for Zn-barlowite).  In these dimerized materials, the hyperfine magnetic field generated by a $^{63}$Cu nuclear spin $\bf{\hat{I}}_\text{i}$ induces singlet-triplet excitation in a pair of Cu electron spins, which in turn induces a hyperfine magnetic field on another $^{63}$Cu nuclear spin $\bf{\hat{I}}_\text{j}$, resulting in a RKKY-like indirect nuclear spin-spin coupling \cite{Ruderman1954} between two  $^{63}$Cu nuclear spins.    In analogy, the oscillatory behavior of $M(2\tau)$ below 50~K indicates that some of the Br sites form a dimerized cluster linked by a Cu-Cu bond, as schematically depicted with light green shade in Fig.~1(b).  We  conducted preliminary spin echo decay measurements at $^{63}$Cu sites \cite{ToBePublished}, and confirmed that $^{63}$Cu also exhibits a damped oscillation with frequency $\omega_\text{o}$ similar to Cu$_2$Sc$_2$Mg$_4$O$_{13}$ (with comparable $J \simeq 260$~K \cite{Kikuchi2013}).  We therefore conclude that the spin echo amplitude oscillations observed below $\sim$50~K reflect the formation of Br-Cu-Cu-Br clusters, which may be related to our prior observation of the gradual emergence of Cu spin singlets below $\sim$30~K \cite{WangNatPhys2021}.

In general, the oscillation of spin echo amplitude can occur only if we use radio frequency pulses to flip a pair of so-called {\it like-spins} resonating at the same frequency \cite{Pennington1989,Pennington1991}; this means that we need to flip a pair of $^{79}$Br-$^{79}$Br nuclear spins, rather than a pair of {\it unlike-spins}, $^{79}$Br - $^{81}$Br.  Since the natural abundance of $^{79}$Br is 51\%, the maximum possible oscillation amplitude is therefore 0.51$M(2\tau=0)$ for $^{79}$Br NQR.  Accordingly, $F\sim0.25$ at 4.2~K implies that the actual fraction of the $^{79}$Br sites involved in the clusters may be as large as $F/0.51 \sim 0.5$.  

It is important to recall, however, that the oscillation of $M(2\tau)$ at low temperatures persists many cycles with little damping  in the case of well-isolated spin dimers in SrCu$_2$(BO$_{3}$)$_2$ \cite{Kodama2002} and  Cu$_2$Sc$_2$Mg$_4$O$_{13}$ \cite{Kikuchi2010}.   On the other hand, $M(2\tau)$ observed for two-leg spin ladders in SrCu$_2$O$_3$ \cite{Ishida1996} and Sr$_{14}$Cu$_{24}$O$_{41}$ \cite{Takigawa1998} exhibits a Gaussian form of decay without oscillations, despite the singlet formation along the rung.  This is because many spin singlets are entangled along the legs, resulting in superposition of many different oscillation frequencies $a_{ij}$, and their average becomes a Gaussian \cite{Pennington1989,Pennington1991}.  In other words, if the spin singlets that emerge below $\sim30$~K in Zn-barlowite \cite{WangNatPhys2021} are isolated in the present case (as in SrCu$_2$(BO$_3$)$_2$ and Cu$_2$Sc$_2$Mg$_4$O$_{13}$), we expect a well defined oscillation with little damping, whereas entanglement of many singlets would lead to a Gaussian (as in SrCu$_2$O$_3$ and Sr$_{14}$Cu$_{24}$O$_{41}$).  The oscillation observed for Zn-barlowite has a clearly defined frequency but with strong damping, and is somewhere between these  two extreme cases.  This underscores the disordered nature of the magnetic ground state in this material.  Note that, theoretically, both nearly isolated and entangled singlets may co-exist within a disordered kagome plane \cite{Kawamura2019}.  A potential caveat of these arguments is that each $^{79}$Br can, in principle, form a large cluster and couple with up to six $^{79}$Br sites within the same interlayer and additional six $^{79}$Br sites in the two adjacent interlayers above and below, as shown in Fig.~1(c).  Simultaneous indirect couplings with many $^{79}$Br sites would cause strong damping in the oscillation.  But diffraction experiments have not detected evidence for such large cluster formation.

To summarize, we used $^{79}$Br NQR to demonstrate that the lattice degrees of freedom in Zn-barlowite undergo gradual freezing below $J\sim 160$~K.    In the frozen state below 50~K, the lattice becomes static with additional structural disorder at local levels.  The oscillation of the spin echo decay induced by indirect nuclear spin-spin interaction indicates that up to $\sim$50\% of Br sites in the frozen state are involved in structural dimer formation encompassing Cu-Cu pairs.  The strong damping of oscillation is inconsistent with completely isolated Cu spin dimers formed in the kagome planes.  On the other hand, a well-defined period of oscillation suggests that Cu spin singlets are not as strongly entangled as in two-leg spin ladders, which exhibit Gaussian decay instead.   The mixed response that we observed is consistent with the notion of closely competitive states in Zn-barlowite that are strongly perturbed by local disorder.

%---------------------------------------------------------------------------
%\begin{acknowledgments}
T.I. thanks K.~Yoshimura for helpful communications, and Y.~Itoh for critical reading of the manuscript.  The work at McMaster was supported by NSERC (T.I.).  P.M.S. was supported by the Rice University Consortium for Processes in Porous Media.  The work at Stanford and SLAC (sample synthesis and characterization) was supported by the U.S. Department of Energy (DOE), Office of Science, Basic Energy Sciences, Materials Sciences and Engineering Division, under contract no. DE-AC02-76SF00515 (Y.S.L. and J.W.).  R.W.S. was supported by a NSF Graduate Research Fellowship (DGE-1656518). 
%\end{acknowledgments}

%\nocite{*}
\bibliographystyle{apsrev4-2}

\newpage
\noindent
{\bf Supplemental Materials for ``Freezing of the Lattice in the Kagome Lattice Heisenberg Antiferromagnet Zn-barlowite ZnCu$_3$(OD)$_6$FBr''}% Force line breaks with \\

\section{\text{$^{79,81}$B\lowercase{r}} and \text{$^{63,65}$C\lowercase{u}} NQR lineshapes}

In Fig.~5, we present the $^{79,81}$Br and  $^{63,65}$Cu NQR lineshapes observed at 4.2~K.  Since $T_2$ is much faster at the $^{63,65}$Cu sites due to the presence of the Cu$^{2+}$ electron spins at the same sites, one can suppress the $^{63,65}$Cu signals by using $\tau=30$~$\mu$s or longer.  This was advantageous for the accurate estimation of the integrated intensity of the $^{79}$Br below 50~K, where $^{63,65}$Cu NQR signals become observable.

\begin{figure}[!h]
\centering
\includegraphics[width=2.8in]{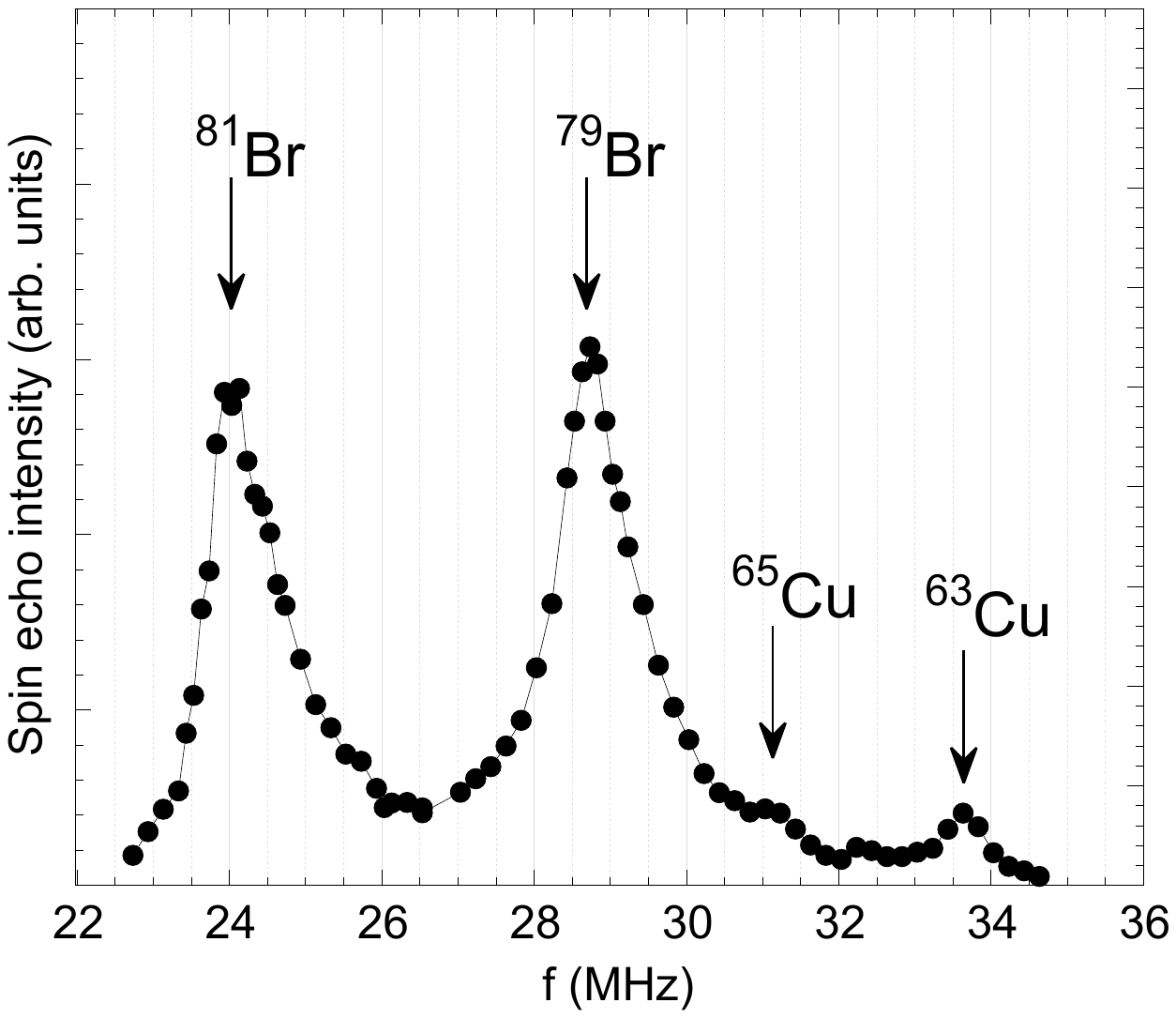}
\caption{$^{79,81}$Br and $^{63,65}$Cu NQR lineshapes observed at 4.2~K.  We measured the lineshape with the pulse separation time  $\tau=30$~$\mu$s (above 26~MHz) or 40~$\mu$s (below 26~MHz) between the 90 and 180 degree radio frequency pulses, and matched the results at 26~MHz.
}
\label{crystal}
\end{figure}

\section{Determination of $1/T_1$}
\subsection{Stretched fit of the $T_1$ recovery curves}
In the interest of simplicity and brevity, we presented the $^{79,81}1/T_{1}$ results determined from the conventional stretched exponential fit of the recovery curve in Fig.~4 and the main text.  In this subsection, we summarize the additional details of the stretched fit results, and the next subsection will be devoted to the more sophisticated inverse Laplace transform (ILT) analysis of the recovery curves, which leads us to the same conclusions.  

In Fig.~6, we summarize the typical recovery curves of the nuclear magnetization $M(t)$ observed at delay time $t$ after an inversion pulse.  The solid curves through the data points represent the empirical stretched fit with 
\begin{align}
M(t) = M_\text{o} - A~exp(-(3t/T_{1})^{\beta}), \label{eq:stretch}
\end{align}  
where the saturated nuclear magnetization $M_\text{o}$, inverted nuclear magnetization $A$, $1/T_1$, and the empirical stretched fit exponent $\beta$ are the fitting parameters.  The pre-factor 3 is for the magnetic relaxation mechanism in the NQR measurements between the nuclear spin $\pm1/2$ and $\pm3/2$ energy levels \cite{Andrew1961,Narath1967}.  The behavior of $\beta$ above 60~K were similar to the results previously reported in Fig.~S2 of \cite{WangNatPhys2021} (we adopted all the results of $^{79,81}1/T_{1}$ below 60~K from \cite{WangNatPhys2021}). 

\begin{figure}[!b]
\centering
\includegraphics[width=3in]{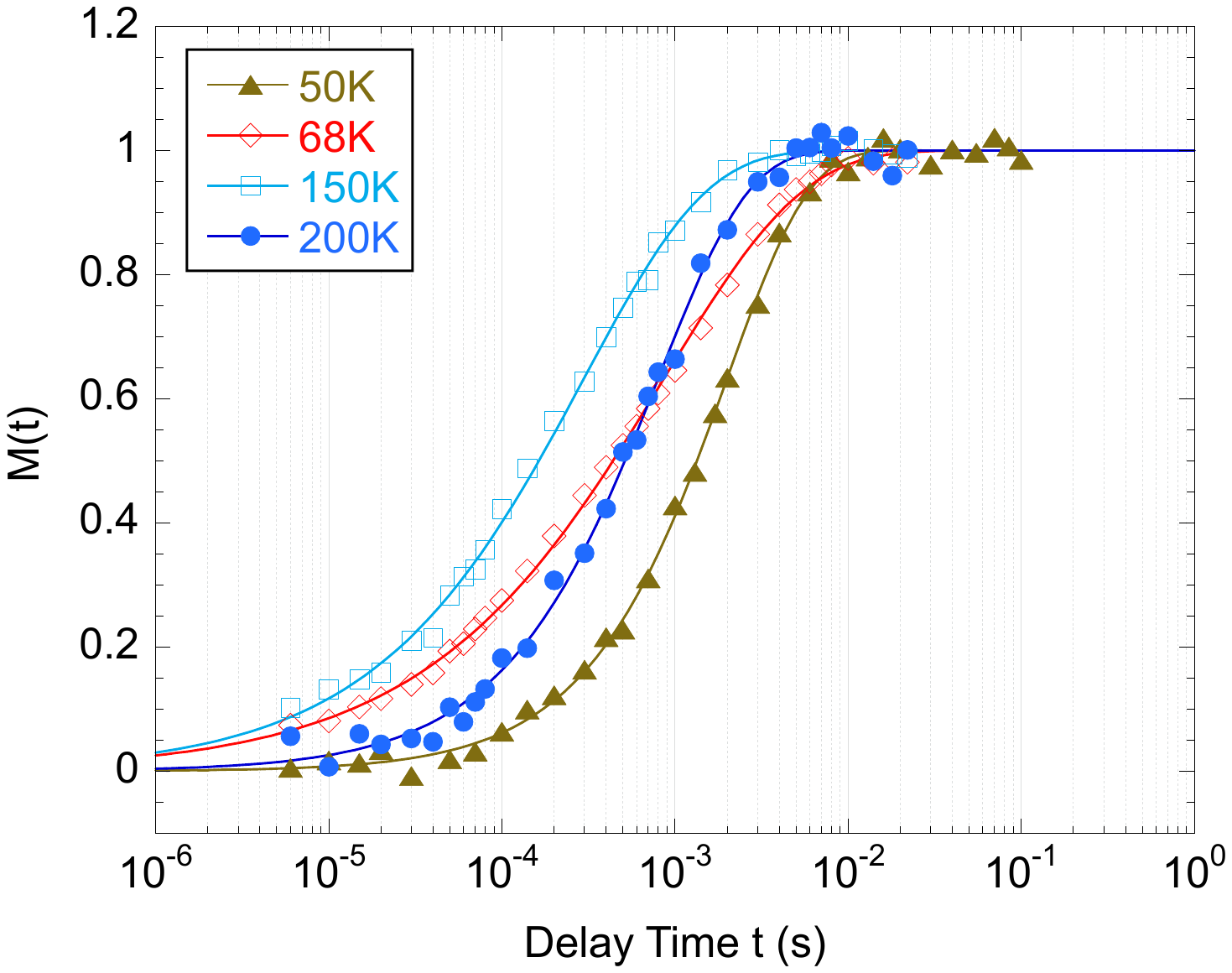}
\caption{Normalized $1/T_{1}$ recovery curves observed for $^{79}$Br at representative temperatures.  The solid curves are the best empirical stretched exponential fit with eq.(1). 
}
\label{crystal}
\end{figure}

From the fit in Fig.~6, we found $\beta = 0.83$ and 0.92 at 200~K and 50~K, respectively.  These values are close to 1, and implies that the distribution in the magnitude of  $^{79}1/T_{1}$ is not very strong.  On the other hand, we found smaller values $\beta = 0.54$ and 0.61 at 68~K and 150~K, respectively.  This is caused by the greater distributions of $^{79}1/T_{1}$ in the intermediate temperature range, where the lattice is freezing.  

The advantage of the conventional stretched fit analysis of the $1/T_1$ recovery curve $M(t)$ is that one can determine the experimental value of $1/T_1$ quite easily, even if the signal to noise ratio is poor.  However, as we repeatedly cautioned in our earlier works on charge ordered high $T_c$ cuprates La$_{1.875}$Ba$_{0.125}$CuO$_4$ \cite{Singer2020} and La$_{1.885}$Sr$_{0.115}$CuO$_4$ \cite{Arsenault2020}, proximate Kitaev spin liquid materials Cu$_2$IrO$_3$ \cite{Takahashi2019} and Ag$_3$LiIr$_2$O$_6$ \cite{WangPRB2021}, as well as Zn-barlowite and herbertsmithite kagome lattice \cite{WangNatPhys2021}, the stretched fit result of $1/T_1$ is only an approximate estimation of $1/T_1$ {\it averaged over the entire sample}.  When $1/T_1$ develops large distributions, it is not always justifiable to rely on the stretched fit, unless the distribution is known to take a certain functional form \cite{JohnstonPRL2005}.  In fact, the closer look at the stretched fit result at 68~K in Fig.~6 suggests that the fit deviates from the data points in a systematic manner.  This is because two distinct components exist in the distribution of $1/T_1$, as shown below based on ILT. 

\subsection{ILTT$_{1}$ analysis}
Regardless of the nature of the distribution, one can calculate the density distribution function  $P(1/T_{1})$ of $1/T_1$ by numerically inverting the experimentally observed $M(t)$ with inverse Laplace transform (ILT) based on Tikhonov regularization,
\begin{align}
M(t) &= \sum_{j=1}^m\left[1-2 \,e^{-3t/T_{1j}}\right]P\!\left(1/T_{1,j}\right). \label{eq:ILTexp}
\end{align}  
Here, $P(1/T_{1,j})$ represents the probability density for a nuclear spin to relax with a particular value of $1/T_{1,j}$.  We refer readers to \cite{Singer2020,mitchell:PNMRS2012} and references therein for the details of ILT.

\begin{figure}
\centering
\includegraphics[width=2.8in]{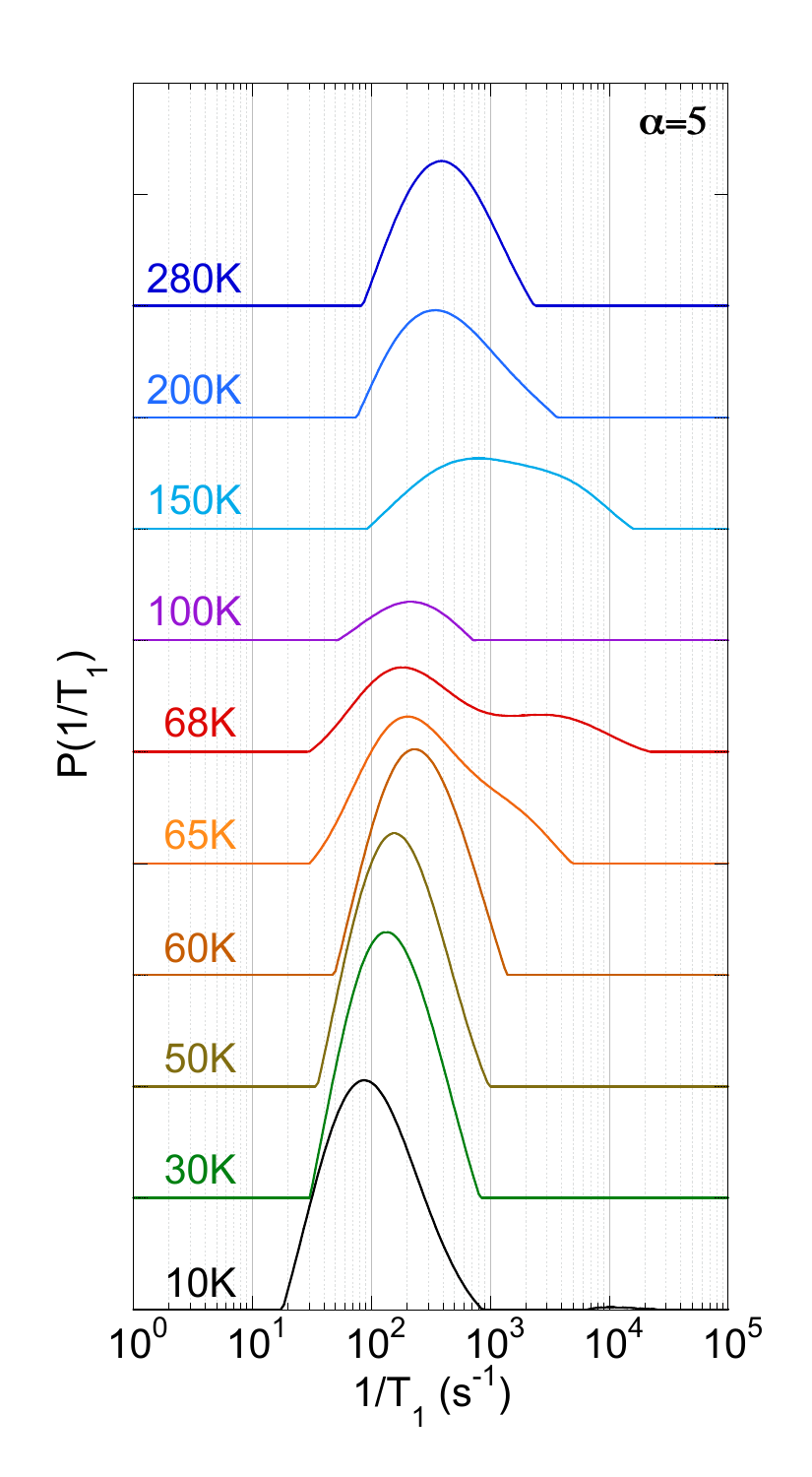}
\caption{Representative density distribution function of $P(1/T_{1})$ for $^{79}$Br.  For clarity, the origin of the vertical axis is shifted at different temperatures.  We fixed the Tikhonov regularization parameter to $\alpha = 5$ to facilitate comparison with equal footing at different temperatures.  The integrated area underneath each curve is normalized to $I_\text{o}$ in Fig.~3(b).
}
\label{crystal}
\end{figure}

In Fig.~7, we present the representative results of $P(1/T_{1})$ calculated from the experimentally observed $M(t)$ based on ILT.  We normalized the integrated area under each $P(1/T_{1})$ curve to the signal intensity $I_\text{o}$ in Fig.~3(b).  Despite up to 48 hours of continuous signal averaging, the signal to noise ratio was limited above 75~K.  Accordingly, we fixed the Tikhonov regularization parameter to a fairly large value $\alpha=5$, implying that $P(1/T_{1})$ is rather strongly smoothed; approximately a half of the total width seen in $P(1/T_{1})$ at 280~K is caused by smoothing of  $P(1/T_{1})$.  

Notice that $1/T_1$ has nearly a symmetrical distribution at 280~K.  But once the lattice freezing sets in below $\sim200$~K, we observed asymmetrical distributions comprising two separate components.  One contribution is always located around $^{79}1/T_{1}^{spin}\simeq 200$~s$^{-1}$, and represents $^{79}$Br nuclear spins relaxing almost entirely due to fluctuating hyperfine magnetic fields from Cu$^{2+}$ electron spins.   Another component manifests itself only in the intermediate temperature range below $\sim200$~K with growing values of $1/T_{1}$ extending up to $\sim10^{4}$~s$^{-1}$.  This component arises from $^{79}$Br nuclear spins under the strong influence of fluctuating EFG, and effectively represented by $^{79}1/T_{1}^{lattice}$ in Fig.~4(c).  We begin to lose the $^{79}$Br NQR signals below 200~K, where some $^{79}$Br nuclear spins relax with  $\sim10^{4}$~s$^{-1}$ or faster.  At 100~K, we can detect only a small fraction of $^{79}$Br nuclear spins, which are still relaxing with $^{79}1/T_{1}^{spin}\simeq 200$~s$^{-1}$, and other $^{79}$Br nuclear spins are not even observable due to strongly enhanced relaxation rates by the EFG.  (For this reason, we did not present the 100~K data point in Fig.~4.)  Upon further cooling, the EFG fluctuations gradually slow down, and increasing numbers of $^{79}$Br nuclear spins become observable again.  At 50~K and below,   $P(1/T_{1})$ regains a symmetrical shape, with no hint of $^{79}$Br nuclear spins with enhanced $^{79}1/T_{1}^{lattice}$.

We emphasize that the rapid relaxation of nuclear spins by lattice fluctuations is  affecting the sample inhomogeneously.  For example, at 68~K, $\sim2/3$ of the observable $^{79}$Br nuclear spins are relaxing with $^{79}1/T_{1}^{spin}\simeq 200$~s$^{-1}$, but other $^{79}$Br nuclear spins are relaxing with $^{79}1/T_{1}^{lattice}\simeq 3000$~s$^{-1}$ or even faster.  This underscores the glassy nature of the freezing of the lattice in Zn-barlowite.  Analogous NMR anomalies of the signal intensity and relaxation rates are often observed in disordered magnetic materials due to inhomogeneous slowing of spin fluctuations \cite{Hunt2001,Arsenault2020,Singer2020}.  The present case is unique, in the sense that these NQR anomalies originate from the slow, inhomogeneous fluctuations of the lattice instead.

%\bibliography{Wang_Lattice_v3a_arXiv}% Produces the bibliography via BibTeX.


%------------------------------------------------------------------------------

\end{document}